\documentstyle[twoside,fleqn,espcrc2,epsf]{article}

\newcommand{\eq}{\begin{equation}}
\newcommand{\en}{\end{equation}}
\newcommand{\eqa}{\begin{eqnarray}}
\newcommand{\ena}{\end{eqnarray}}

\newcommand{\tr}{\mbox{tr}\,}
\newcommand{\AmS}{{\protect\the\textfont2
  A\kern-.1667em\lower.5ex\hbox{M}\kern-.125emS}}

\title{Casimir scaling or flux counting~?
\vskip -1truecm\rightline{HUB-EP-99/44}\vskip 1truecm}

\author{Gunnar~S.~Bali\thanks{Talk presented
at XVIIth International Symposium on Lattice Field Theory (Lattice 99),
Pisa, 28.6.\ -- 3.7.1999.}\address{Institut f\"ur Physik, Humboldt Universit\"at
zu Berlin,
Invalidenstr.\ 110, 10115 Berlin, Germany}}
      
\begin{document}

\begin{abstract}
Potentials between two static sources in various
representations of the SU(3) gauge group are determined
on anisotropic $3+1$ dimensional lattices. Strong evidence
in favour of ``Casimir scaling'' is found.
\end{abstract}

\maketitle

\section{THE QUESTION}
\noindent
Static colour sources offer an ideal environment for investigating
the confinement mechanism and testing models of low energy QCD.
Despite the availability of
a wealth of information on fundamental potentials~\cite{bali1,bali},
only few lattice investigations of forces 
between sources in higher
representations of the gauge group SU(N) exist.
Most of these studies have been done in SU(2) gauge theory in
three~\cite{3dsu2,3dsu22} and four~\cite{4dsu2} space-time dimensions.
Only two groups have obtained results for
four dimensional SU(3) gauge theory so far, one of them
at finite temperature from Polyakov lines~\cite{polya} and the other
from Wilson loops~\cite{su3,michael} at zero temperature.

For the static potential in the singlet channel, tree level
perturbation theory yields the result,
\begin{equation}
V_D(r)=-C_D\frac{g^2}{4\pi}\frac{1}{r},
\end{equation}
where $D=1,3,6,8,10,\ldots$ labels the representation of SU(3).
$D=3$ corresponds to the fundamental representation, $F$, and
$D=8$ to the adjoint representation, $A$. $C_D$ labels the corresponding
quadratic Casimir operator $C_D=\mbox{Tr}_Dt_a^Dt_a^D$ with
the trace $\mbox{Tr}_D$ and generators $t^D_a$ obeying the normalisation
conditions, $\mbox{Tr}_D {\mathbf 1}_D = 1$, $[t_a^D,t_b^D]=if_{abc}t_c^D$.
The Table below contains all representations $D$ that we have realised,
the corresponding weights $(p,q)$ and
the ratios of Casimir factors, $d_D=C_D/C_F$. In SU(3) we have $C_F=4/3$
and $z=\exp(2\pi i/3)$ denotes a third root of {\em one}.
\begin{table}[hbt]
\setlength{\tabcolsep}{0.95pc}
\label{tab1}
\begin{tabular}{crccl}
\hline
$D$&$(p,q)$&$z^{p-q}$&$p+q$&$d_D$\\\hline
3&$(1,0)$&$z$&1&1\\
8&$(1,1)$&1&2&2.25\\
6&$(2,0)$&$z^*$&2&2.5\\
$15a$&$(2,1)$&$z$&3&4\\
10&$(3,0)$&1&3&4.5\\
27&$(2,2)$&1&4&6\\
24&$(3,1)$&$z^*$&4&6.25\\
$15s$&$(4,0)$&$z$&4&7\\
\hline
\end{tabular}
\vskip -.7cm
\end{table}

We denote group elements in the fundamental
representation of SU(3) by $U$. 
The traces of $U$ in various representations,
$T_D=\tr U_D$, can easily be obtained,
\begin{eqnarray}
T_3&=&\tr U,\quad T_8=\left(|T_3|^2-1\right),
\nonumber\\
T_6&=&\frac{1}{2}\left[(\tr U)^2+\tr U^2\right],
\nonumber\\\label{eq:reps}
T_{15a}&=&\tr U^*\,T_6-\tr U,\\
T_{10}&=&\frac{1}{6}\left[(\tr U)^3+3\,\tr U\,\tr U^2+2\,\tr U^3\right],\nonumber\\
T_{24}&=&\tr U^*\,T_{10}-T_6,\quad
T_{27}=|T_6|^2-|T_3|^2,\nonumber\\
T_{15s}&=&\frac{1}{24}\left[
(\tr U)^4+6(\tr U)^2\tr U^2\right.\nonumber\\\nonumber
&+&\left.3(\tr U^2)^2+8\,\tr U\,\tr U^3+6\,\tr U^4\right].
\end{eqnarray}
Note the difference, $\mbox{Tr}_DU_D=\frac{1}{D}\tr U_D=T_D/D$.
Under the replacement $U\rightarrow zU$, $T_D\rightarrow z^{p-q}T_D$.

It is known~\cite{bali1,bali} that
for distances $r=R/a_s\geq 0.6\,r_0\approx$~0.3~fm the
fundamental potential is well described by the
parametrisation,
\begin{equation}
\label{eq:param}
V_F(R)=V_{0,F}-\frac{e_F}{R}+K_FR.
\end{equation}
Perturbation theory tells us,
$V_{0,R}\approx d_RV_{0,F}$, $e_R\approx d_Re_F$.
While the fundamental potential in pure gauge theories linearly
rises {\em ad infinitum}, the adjoint potential
will be screened by gluons and,
at sufficiently large distance, decay into two gluelumps (or gluinoballs),
bound states of a static adjoint source
(gluino) and gluons.
This string breaking has indeed been confirmed~\cite{3dsu22}. Therefore,
strictly speaking, the adjoint string tension is {\em zero}. 
In fact, all charges in higher than the fundamental representation
will be at least partially screened by the background gluons. For instance,
$6\otimes 8=24\oplus 15a^*\oplus 6\oplus 3^*$:
in interacting with the glue, the sextet potential obtains a fundamental
component. However,
in an intermediate range an approximate linear behaviour is
found~\cite{3dsu2,3dsu22,4dsu2,polya,su3,michael}, such that
one might speculate
whether in this region
the Casimir scaling hypothesis $K_D\approx d_DK_F$ holds.

A simple rule, related to the centre of
the group, is reflected in Eq.~(\ref{eq:reps}):
where ever $z^{p-q}=1$, the source will be reduced into a singlet component
at large distance while, where ever
$z^{p-q}=z (z^*)$, it will be screened up to a residual (anti-)\-triplet
component, i.e.\
one can easily read off the asymptotic string tension ($0$ or $K_F$) from
the third column of the Table, rather than having to multiply
and reduce representations. If centre symmetry plays such a
prominent r\^ole at infinite distance, one might imagine
the intermediate distance slopes
to count the number of fundamental flux tubes
$p+q$ that are embedded into the higher representation vortex.
This flux counting model is supported
by indications that the QCD vacuum lies on the boundary between a type I and
a type II superconductor~\cite{bali2} and, therefore,
interactions between neighbouring vortices
are weak. Indeed, latest results
on the adjoint SU(2) potential~\cite{3dsu22} as well as on
the octet and sextet SU(3) potentials~\cite{michael}
turn out to lie significantly lower than the expected Casimir ratios suggest.
Expectations from both models,
Casimir scaling and flux counting, at least for the lower
dimensional representations, are close to each
other, such that discriminating between them
represents a numerical challenge.

\section{THE METHOD}
\begin{figure}[htb]
\leavevmode
\epsfxsize=7.5cm\epsfbox{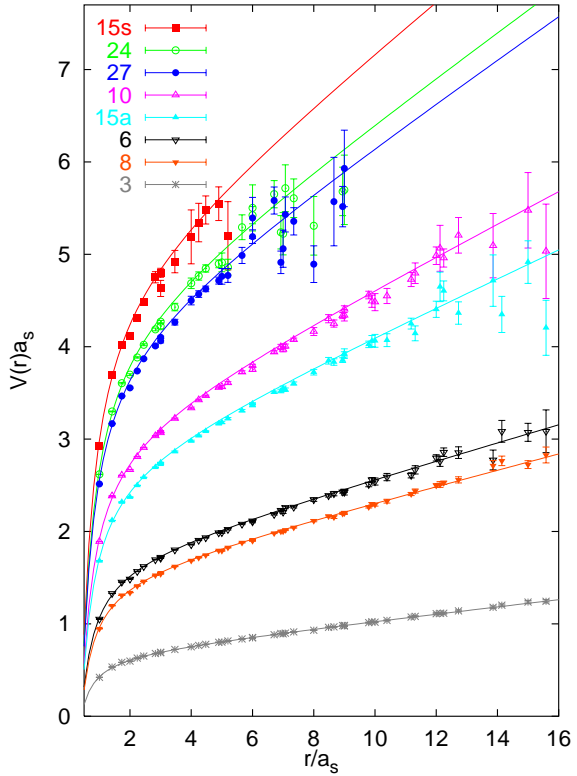}
\vskip -1.cm
\caption{Static potentials between sources in various representations
of SU(3).\vskip -.8cm}
\label{fig1}
\end{figure}
\noindent
We simulate SU(3) gauge theory on anisotropic lattices,
using the standard plaquette action,
\begin{equation}
S=\beta\left(\frac{1}{\xi_0}\sum_{x,i>j}S_{x,ij}+\xi_0\sum_{x,i}S_{x,i4}\right).
\end{equation}
We determine the renormalised anisotropy
from the (fundamental)
spatial potential, $a_sV_s(R_sa_s)=a_sV_s(R_ta_t)=a_sV_s(0.8\,r_0)$.
For the combination, $\beta=6.2$ and $\xi_0=3.25$, we find $\xi=a_s/a_t=
R_t/R_s\approx 4$,
$r_0\approx 6\,a_s$, i.e.\ ${a_t}^{-1}\approx 9.6$~GeV,
$a_s^{-1}\approx 2.4$~GeV.
We present only preliminary results from
an analysis of a subset of our final statistics at these parameter values.
Additional simulations are being performed at courser lattices
with constant renormalised anisotropy.

The potentials are extracted in the standard way from fits
to spatially smeared Wilson loops~\cite{bali1}.
Wilson loops in higher representations
are obtained from the fundamental one by use of Eq.~(\ref{eq:reps}).
In SU(3), unlike in SU(2)~\cite{3dsu22}, this procedure excludes
thermal averaging of temporal links.

\section{THE ANSWER}
\noindent
In Fig.~\ref{fig1}, we display our results in lattice units,
$a_s\approx 4\,a_t$.
$r=12\,a_s$ corresponds to a distance $r\approx 1$~fm. Note, that
we display the raw lattice data and have not subtracted any self energy piece.
We fit the fundamental potential for distances $r\geq 0.6~r_0$
to Eq.~(\ref{eq:param}). The expectations on the potentials $V_D(r)$,
which are dispayed in the Figure, correspond to this fit curve,
multiplied by the factors $d_D$.
As one can see, up to distances where the signal disappears into noise or the
string might break, the data is well described by the
Casimir scaling assumption.

\noindent
\begin{figure}[htb]
\leavevmode
\vskip -.75cm
\epsfxsize=7.5cm\epsfbox{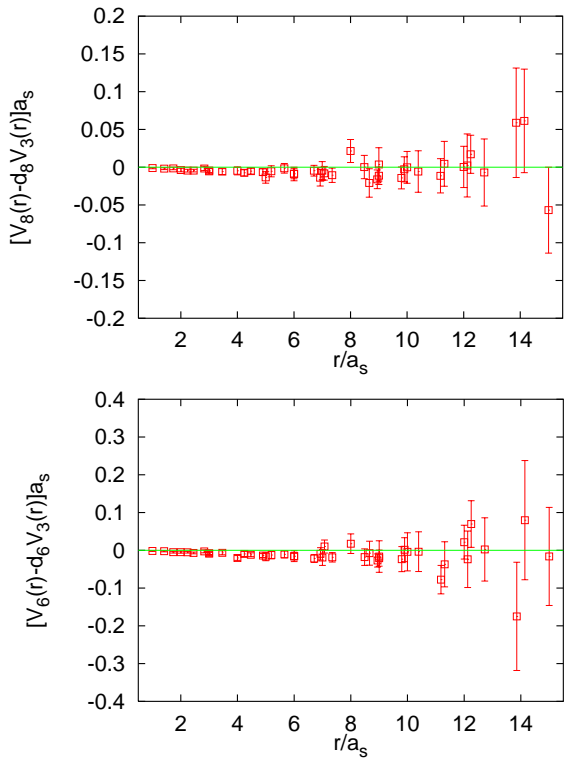}
\vskip -.8cm
\caption{Test of the Casimir scaling hypothesis.
\vskip -.5cm}
\label{fig2}
\end{figure}

In Fig.~\ref{fig2}, we have subtracted the fundamental potential, multiplied
by the factors $d_8$ and $d_6$ from the adjoint and sextet potentials,
respectively. We benefit from a reduction of statistical errors,
due to correlations between the data sets. While the flux
counting model is certainly ruled out, the Casimir scaling model tends
to slightly overestimate
the data points. 
For higher representations,
the situation looks similar.
Compared to results from coarser lattices~\cite{michael}
($r_0\approx 3\,a$ as opposed to $r_0\approx 24\,a_t$) the disagreement, however,
is vastly reduced and will eventually completely disappear
in the continuum limit.
A comprehensive study with proper extrapolation to the continuum limit is
in preparation.

\section*{ACKNOWLEDGEMENTS}
\noindent
This research has been funded funding by the Deutsche Forschungsgemeinschaft
(grants Ba~1564/3-1 and Ba~1564/3-2).
Computations were performed on the Cray T90 of
the Neumann-Institut for Computing in FZ J\"ulich.

\end{document}